\documentclass[sigconf]{acmart}

\usepackage{multirow}
\AtBeginDocument{%
  \providecommand\BibTeX{{%
    \normalfont B\kern-0.5em{\scshape i\kern-0.25em b}\kern-0.8em\TeX}}}

\setcopyright{acmcopyright}
\copyrightyear{2021}
\acmYear{2021}
\acmConference[ICPP'21]{The 50th International Conference on Parallel Processing}{August 9--12, 2021}{Chicago (Argonne National Lab), Illinois, USA}
\acmBooktitle{The 50th International Conference on Parallel Processing, August 9--12, 2021, Chicago (Argonne National Lab), Illinois, USA}
\begin{document}

% \title{MetaCache GPU - }
\title{MetaCache-GPU: Ultra-Fast Metagenomic Classification}
%================================================================================

\author{Robin Kobus}
\email{kobus@uni-mainz.de}
\affiliation{%
  \institution{Institute of Computer Science\\ {Johannes Gutenberg University}}
  \city{Mainz}
  \country{Germany}
}

\author{Andr\'{e} Müller}
\email{muellan@uni-mainz.de}
\affiliation{%
  \institution{Institute of Computer Science\\ {Johannes Gutenberg University}}
  \city{Mainz}
  \country{Germany}
}

\author{Daniel Jünger}
\email{juenger@uni-mainz.de}
\affiliation{%
  \institution{Institute of Computer Science\\ {Johannes Gutenberg University}}
  \city{Mainz}
  \country{Germany}
}

\author{Christian Hundt}
\email{chundt@nvidia.com}
\affiliation{%
 \institution{NVIDIA AI Technology Center}
 \city{Luxembourg}
 \country{Luxembourg}
 }

\author{Bertil Schmidt}
\email{bertil.schmidt@uni-mainz.de}
\affiliation{%
  \institution{Institute of Computer Science\\ {Johannes Gutenberg University}}
  \city{Mainz}
  \country{Germany}
}

\renewcommand{\shortauthors}{Kobus, et al.}

\begin{abstract}

The cost of DNA sequencing has dropped exponentially over the past decade, making genomic data accessible to a growing number of scientists. In bioinformatics, localization of short DNA sequences (reads) within large genomic sequences is commonly facilitated by constructing index data structures which allow for efficient querying of substrings.
Recent metagenomic classification pipelines annotate reads with taxonomic labels by analyzing their $k$-mer histograms with respect to a reference genome database. CPU-based index construction is often performed in a preprocessing phase due to the relatively high cost of building irregular data structures such as hash maps. However, the rapidly growing amount of available reference genomes establishes the need for index construction and querying at interactive speeds.
In this paper, we introduce MetaCache-GPU -- an ultra-fast metagenomic short read classifier specifically tailored to fit the characteristics of CUDA-enabled accelerators. Our approach employs a novel hash table variant featuring efficient minhash fingerprinting of reads for locality-sensitive hashing and their rapid insertion using warp-aggregated operations.
Our performance evaluation shows that MetaCache-GPU is able to build large reference databases in a matter of seconds, enabling instantaneous operability, while popular CPU-based tools such as Kraken2 require over an hour for index construction on the same data.
In the context of an ever-growing number of reference genomes, MetaCache-GPU is the first metagenomic classifier that makes analysis pipelines with on-demand composition of large-scale reference genome sets practical.
The source code is publicly available at \url{https://github.com/muellan/metacache}.

%A common paradigm in bioinformatics is to store and index sequence data as sets of $k$-length substrings (called $k$-mers).
%\cite{marchet2019data}.
%We have thus explored the efficiency of WC for indexing large amounts of genomes for metagenomic classification tasks in comparison to the popular CPU-based tools Kraken2~\cite{Kraken2} and MetaCache~\cite{muller2017metacache} -- both using hash tables as their primary index data structure -- as a case study.

\end{abstract}




\maketitle

%================================================================================
\section{Introduction}
%================================================================================

Recent years have seen a tremendous increase in the volume of data generated in the life sciences, especially propelled by the rapid progress of {\em next generation sequencing} (NGS) technologies \cite{stephens2015big}. High-throughput sequencers can produce massive amounts of short DNA strings (called reads) in a single run. This leads to large-scale datasets being processed in a wide range of bioinformatics applications.
%including de-novo sequencing, re-sequencing, metagenomics, transcriptomics, and epigenetics.
Furthermore, the cost of these technologies has been decreasing dramatically. The low sequencing cost per genome\footnote{Currently around US\$1,000 per genome: see \url{http://www.genome.gov/sequencingcosts}} enables widespread usage and renders population-scale projects feasible.
%with increasing statistical power provided by larger data sets.
Examples include
%the sequencing of the whole population of Iceland \cite{gudbjartsson2015large},
the Earth BioGenome Project \cite{lewin2018earth}, metagenomics microbiome sequencing studies \cite{korpela2016intestinal}, and world-wide SARS-CoV-2 sequencing efforts \cite{bauer2020supporting}.

However, the analysis of large sequencing datasets poses hard computational challenges. In particular, read mapping and read classification are per\-for\-mance-critical tasks, being initial stages required for manifold types of NGS analysis pipelines. The objective of read mapping is to determine the best mapping location(s) of each read in a given (set of) reference genome(s). A common paradigm to address this issue is to store and index reference genome sequences as sets of $k$-length substrings (called $k$-mers) \cite{marchet2019data}. The constructed index is queried for the $k$-mers in each read to find exact matches, which are further processed using a seed-and-extend approach.

Nevertheless, associated runtimes on common workstations remain high when processing reads at scale. Parallelization can be employed to reduce execution times but imposes additional challenges due to variable sizes of $k$-mer matches, large storage requirements of index data structures, and their associated irregular memory access patterns \cite{Reinert15}. As a consequence, corresponding speedups on accelerators such as GPUs and FPGAs are often limited \cite{houtgast2018hardware, houtgast2015fpga}. Aforementioned difficulties are amplified in {\em metagenomic} read classification where a large number of reference genomes is considered; e.g., the recent NCBI RefSeq Release 202 contains 51,326 genomic sequences from 15,461 different species. Metagenomic read classification thus requires both the rapid construction and high throughput querying of large index data structures, since reference genome collections are subject to frequent change and continuous growth.

In summary, algorithmic design and implementations of bioinformatics applications struggle to keep pace with recent high-throughput sequencing techniques and their ever increasing data acquisition rates. In this paper, we present MetaCache-GPU -- a metagenomic read classification algorithm optimized for CUDA-enabled accelerators. We demonstrate how to efficiently construct and query a $k$-mer index for large genome collections employing a novel massively parallel multi-bucket hash table. Furthermore, we leverage multiple GPUs to overcome storage limitations of the scarce video memory of a single GPU. The combination of both approaches leads to up to 69 and 72 times faster database builds and
up to 153 and 6 times
% 3 to 153 and 2 to 6 times
faster queries compared to the established CPU-based applications MetaCache \cite{muller2017metacache} and Kraken2 \cite{wood2019kraken2}, respectively.
Furthermore, MetaCache-GPU's {\em on-the-fly} mode avoids saving and reloading the database, enabling nearly instantaneous operability. Thus, it allows for querying the database directly after construction and is up to 410 and 450 times faster than MetaCache and Kraken2, respectively.

The main contributions of this paper are four-fold.
\begin{itemize}
    \item A novel multi-value hash table that enables rapid and memory-frugal construction and querying of $k$-mer indices on GPUs
    \item The corresponding GPU-based minhashing scheme and top candidate generation for data-parallel read classification
    \item In-memory index construction that allows for on-the-fly classification pipelines avoiding intermediate disk I/O
    \item Index distribution across multiple GPUs to support reference genome $k$-mer indices exceeding single GPU memory
\end{itemize}

The rest of this paper is organized as follows. Section \ref{sec:background} provides necessary background on the topic of metagenomic classification. Related work is discussed in Section \ref{sec:related}. The general MetaCache pipeline is described in Section \ref{sec:cpu}. The design and implementation of our proposed GPU-based MetaCache pipeline are detailed in Section \ref{sec:gpu:impl}. Section \ref{sec:evaluation} evaluates performance and Section \ref{sec:conclusion} concludes.

%================================================================================
\section{Background}\label{sec:background}
%================================================================================
The analysis of the taxonomic composition of a sequenced environmental sample is a fundamental building block in metagenomic pipelines. The corresponding classification problem aims at assigning a suitable taxonomic label (e.g., a species or a genus) to a given NGS read. A traditional approach addressing this problem aligns each read to an annotated database of reference genome sequences.

More concretely, consider a collection of reference (genome) sequences $G = \{G_1,\ldots,G_n\}$ and a collection of short reads $R = \{R_1,\ldots,R_m\}$ with sequence lengths $\left|  G_i \right| >> \left|  R_j \right|$, $\forall  i \in \{ 1,\ldots,n \}$, $\forall j \in \{ 1,\ldots,m \}$. The objective of {\em metagenomic classification} is to identify the most likely genome in $G$ that each read in $R$ may originate from. Note that exact matches of a complete read to a reference substring is unlikely due to genomic variation and sequencing errors. Thus, partial and inexact matches must be considered. However, the corresponding measures can be compute-heavy; e.g., calculating the optimal semi-global alignment score with commonly used dynamic programming algorithms between a read $R_j$ and a reference $G_i$ exhibits a time complexity proportional to the product of sequence lengths; i.e., $\mathcal{O}(\left|  G_i \right| \cdot \left|  R_j \right|)$. Since $n \cdot m$ such alignments must be computed, corresponding runtimes would be prohibitively long.

More recent alignment-free tools can reduce the high complexity based on exact $k$-mer matching. In this approach, a $k$-mer index data structure (or database) is constructed in a preprocessing step. The index is usually based on a hash table that contains all distinct substrings of length {\em k} of each reference in $G$ as keys and their corresponding locations as values. A read $R_j$ is then classified by a look-up procedure, which extracts the set of all $k$-mers in $R_j$ and subsequently queries it against the precomputed index. If the look-up returns matches, counters for the corresponding reference genomes are incremented. At the end of this procedure $R_j$ can be classified based on high-scoring counters. Kraken \cite{kraken} is a highly popular metagenomic classification tool following this approach.

However, the number of sequenced genomes is rapidly increasing. Thus, the index data structure storing the $k$-mers of each reference sequence can become exceedingly large. MetaCache \cite{muller2017metacache} addresses this problem by applying a distinct subsampling technique called {\em minhashing} \cite{Broder2000}. A minhashing filter with sketch size $s$ selects those $k$-mers within a sequence $G$ for a sketch $S(G)$ whose hash values are among the $s$ smallest. A simple example using $k=4$ and $s=2$ looks as follows:
\begin{small}
\begin{verbatim}
    sequence:   ACTGACTG
    4-mers:     ACTG      hash(ACTG) = 14
                 CTGA     hash(CTGA) =  8  <- select
                  TGAC    hash(TGAC) =  7  <- select
                   GACT   hash(GACT) = 11
                    ACTG  hash(ACTG) = 14
\end{verbatim}
\end{small}

When comparing two sequences which differ significantly in their length, as it is the case when comparing short reads to reference genomes, it is advantageous to construct a locality sensitive sketch representation which is constrained to a certain region (window) of the reference. Note that the Jaccard index of two sketches approximates the true Jaccard index of the whole sets of $k$-mers of the corresponding windows and hence can be used as proxy.

%================================================================================
\section{Related Work}\label{sec:related}
%================================================================================

Wood and Salzberg \cite{kraken} were among the first to demonstrate that a $k$-mer-based exact matching approach can achieve high metagenomic classification accuracy by developing Kraken while being around three orders-of-magnitude faster than the alignment tool MegaBLAST \cite{MegaBLAST}. Recent benchmark studies \cite{lindgreen, lemmi} demonstrated that $k$-mer based tools such as Kraken \cite{kraken}, Kraken2 \cite{wood2019kraken2}, CLARK \cite{clark}, and MetaCache \cite{muller2017metacache} can produce superior read assignment accuracy for selected bacterial metagenomic datasets. While being accurate, the major drawback of the $k$-mer based approach is high main memory consumption and long database construction times. For medium-sized bacterial reference genome sets the databases used by Kraken and CLARK already consume several hundreds of gigabytes in size. MetaCache and Kraken2 reduce the $k$-mer index memory consumption by around one order-of-magnitude by using different $k$-mer subsampling techniques (minhashing resp. minimizers). Nevertheless, the rapidly increasing amount of available bacterial reference genomes and the significantly higher complexities of eukaryotic reference genomes relevant for applications such as the monitoring of food ingredients \cite{kobus2020big} demand scalable solutions that can build and query $k$-mer indices at higher speed than current approaches. In this paper, we investigate how modern multi-GPU systems can be used to accelerate MetaCache. Note that the GPU-based $k$-mer index structure introduced in this paper can be easily applied to related bioinformatics tasks such as read mapping \cite{Reinert15} or long-read-to-long-read alignment \cite{ellis2019dibella}.

There has been a limited amount of prior work on using GPUs for metagenomic read classification. cuCLARK \cite{kobus2017accelerating} accelerates CLARK using CUDA but only achieves speedups between 3.2 and 6.6 while keeping the high memory consumption of CLARK. MetaBinG2 \cite{qiao2018metabing2} applies a hidden Markov model to estimate the distance of a read to organisms but is over an order-of-magnitude slower compared to CLARK. Usage of $k$-mer indices on GPUs has also been investigated for the related problem of aligning (or mapping) reads to a single genome. However, most approaches are based on accelerating popular short read aligners such as BWA-MEM or Bowtie2 adopting the FM-index and Burrow-Wheeler-Transform (BWT) \cite{liu2013cushaw2, chacon2014boosting}. While this data structure can be memory efficient, $k$-mer querying requires iterative lookups with irregular memory accesses. Thus, reported speedups are relatively low; e.g. \cite{houtgast2018hardware} only reports a speedup of 2 on a GPU compared to CPU-based BWA-MEM. None of these papers consider the acceleration of $k$-mer index construction.
$k$-mer histogram generation is another task that relies on such $k$-mer index data structures and has been studied
%extensively for CPUs \cite{marcais2012jellyfish}, CPU clusters \cite{
%kokot2017kmc, rizk2013dsk,
%pan2018optimizing} and
for GPUs \cite{li2018gpu, cadenelli2017accelerating, erbert2017gerbil}. However, counting $k$-mer occurrences is a far simpler task compared to the construction of a reference index, i.e. a multi-value key-value store, which is needed for metagenomic classification. Distributed $k$-mer hash tables have also been studied on CPU clusters for $k$-mer counting \cite{pan2018optimizing} and de-novo assembly \cite{georganas2018extreme}, while our work focuses on using multi-GPU systes for metagenomics.

In this paper, we introduce a novel multi-value hash table variant optimized for memory-efficient $k$-mer index construction and querying on multiple GPUs. Database (or $k$-mer index) construction performance is predominantly governed by the throughput of the underlying hash table implementation. To alleviate this bottleneck, we leverage the  fast memory interface of modern CUDA accelerators. High-throughput GPU hash tables have been studied extensively \cite{lessley2019data}.
However, most existing implementation show limitations which make them unsuitable for our use case.
In particular none of the implementations feature out-of-the-box multi-GPU support which is key for metagenomic classification since many real world databases exceed the memory space of a single GPU.
As a more recent publication, WarpCore \cite{warpcore}, successor of WarpDrive \cite{junger2018warpdrive}, proposes a framework that allows for the design of purpose-built GPU hash tables that can be tailored towards optimal performance for a given use case and can
outperform previous approaches such as cuDPP \cite{Alcantara2011} and SlabHash \cite{ashkiani2018dynamic}.
Their cooperative probing scheme uses sub-warp tiles, i.e., CUDA cooperative groups, over a hybrid two-stage probing scheme, where an outer double hashing strategy is used to suppress table clustering effects, while an inner group-parallel linear probing scheme ensures coalesced memory access.
Additionally, the authors propose a multi-GPU extension based on an efficient all-to-all communication pattern over dense NVLink topologies \cite{kobus2019gossip}.
In this work we extend the aforementioned WarpCore framework by a novel hash table layout which better suits typical $k$-mer distributions in terms of performance as well as storage density.

Another time-consuming step in our GPU pipeline is sorting lists of target locations resulting from database queries. A segmented sort algorithm can be employed to efficiently sort multiple lists in a batch of results. While CUDA libraries like CUB \cite{cub}, CUSP \cite{cusp} and ModernGPU \cite{moderngpu} provide segmented sort implementations, the implementation by Hou et al. \cite{Segsort} has been shown to outperform all of the above. Their segmented sort consists of multiple kernels, each tailored to a specific range of segment sizes. In each CUDA kernel, threads operate conjointly sorting the elements of a segment using bitonic sort. The corresponding sorting network can be implemented efficiently exploiting fast register accesses and warp shuffles. We adapt this approach to allow for key-only sorting. In addition, our implementation reduces memory overhead and features asynchronous execution of all invoked kernels.

\section{MetaCache Pipeline}\label{sec:cpu}
%================================================================================

%****************************************************
% one figure for cpu w/o partition and another for multi-gpu ?
	\begin{figure*}[t]
	    \includegraphics[width=0.79\textwidth]{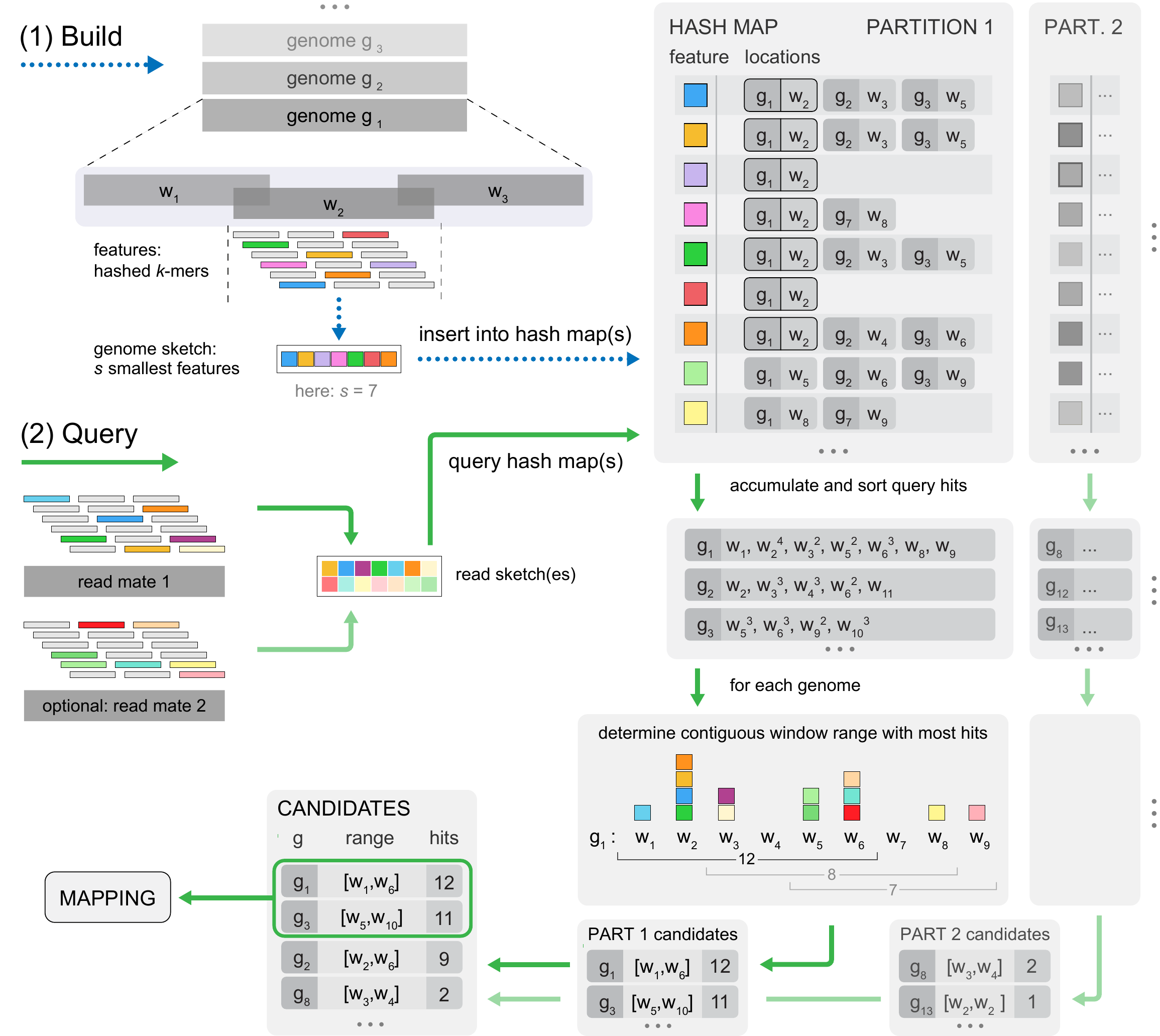}
		\caption{Workflow:
	    (1) Database construction: Each reference genome $g_i$ is partitioned into slightly overlapping windows $w_j$.
		The $s$ smallest $k$-mer hashes of each window are inserted into the database.
		(2) Classification: a database is queried with the $s$ smallest $k$-mer hashes of a read (or read pair). Subsequent counting of hits within each window results in a list of mapping candidates.
		In the case of several database partitions, the candidates from distinct partition are merged in order to assign a read to the most likely taxon of origin.}
		\label{fig:database_construction}
	\end{figure*}
%****************************************************

    % 	The hash table (database) for a given collection of reference genomes is constructed using open addressing. The entries of the hash table consist of key-target-list pairs. An associated hash function $h_2$ maps $k$-mers to slots in the hash table. If an identified slot is empty or occupied with the same $k$-mer, the corresponding $k$-mer is inserted as key and the corresponding location (genome ID, window ID) is appended to the target-list. If the slot is occupied by a different $k$-mer quadratic probing is used to iterate to the next slot. Target lists have a pre-defined maximum length. If the maximum length is reached, the corresponding $k$-mer is considered uninformative and deleted from the hash table at the end of the construction.

    The workflow of MetaCache can be separated into two distinct phases: {\em build} and {\em query}. First, a reference database must be build. Then, reads from metagenomic samples are classified after querying the database. Typical for index based methods, the default MetaCache pipeline is split into separate program calls for build and query, where the database has to be saved to and loaded from disk.
    Here, querying can be executed in different modes, either a single run processing all supplied input files or an interactive session, which holds the database in memory and allows for performing an arbitrary number of queries in succession.
    Additionally, we extended MetaCache with a novel on-the-fly mode where queries can be executed directly after building the database before writing it to the file system. To the best of our knowledge, lightweight in-memory queries are unprecedented in the literature since index construction times have been prohibitively long for competing, non-GPU-based solutions.
    Figure \ref{fig:database_construction} shows an overview of the build and query phases which are further described in detail in the following.

	\begin{figure*}[t]
	    \includegraphics[width=0.9\textwidth]{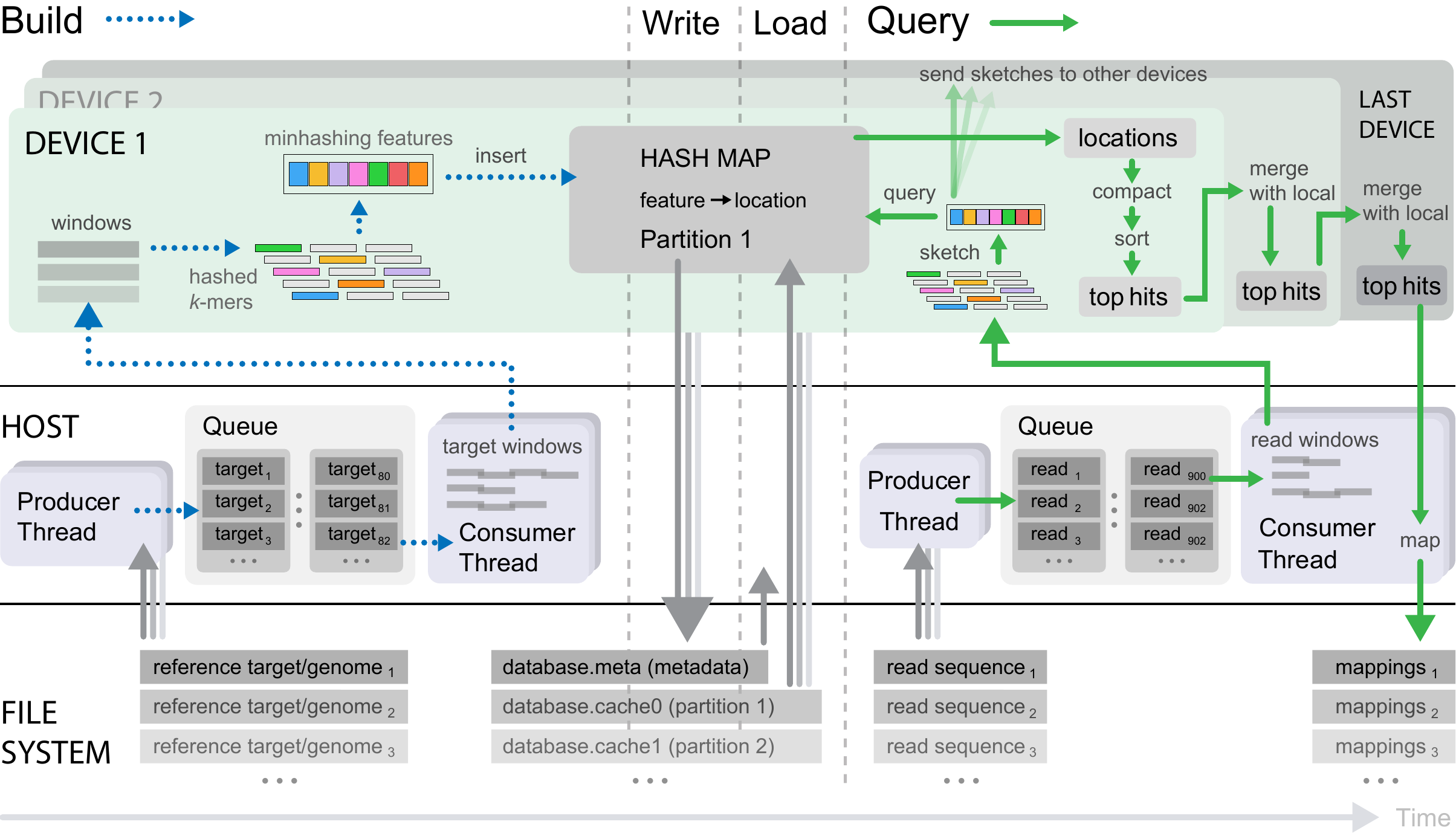}
		\caption{GPU Workflow:  (1) Build:
		producer threads enqueue genomic sequences. Consumer threads split them into windows and copy these to the GPUs. GPUs generate minhashing sketches and insert them into their local hash maps.
		(2) Query: batches of read windows are copied to the first GPU for sketch generation.
		Each GPU queries its local hash map and sends the sketches to the next GPU.
		Resulting location lists are compacted and sorted. Top locations are selected and sent
		to the next GPU for merging. The last GPU obtains the final top list and sends it to the host for read mapping and output. }
		\label{fig:gpu_pipeline}
	\end{figure*}

\subsection{Build Phase} \label{sec:cpu:build}

    Building a reference database starts by generating a taxonomic tree containing the relations of the considered reference genomes. The nodes and structure of the tree are extracted from files provided by NCBI \cite{ncbitax}.
    Next, a hash table is constructed that maps $k$-mers (keys) to genome locations (values). Here, MetaCache employs a producer-consumer strategy to fill a concurrent queue with batches of reference sequences. Multiple producer threads parse the genome files to split the data into header and sequence strings which are then pushed into the queue.
    A consumer thread dequeues a batch of sequences and processes one sequence after another. First it obtains the genomic identifier from the header to create a reference target which is then connected to the taxonomic tree structure. Next, it proceeds to process the corresponding sequence data by dividing it into windows of size $w$ overlapping by $k-1$ base-pairs for a $k$-mer length of $k$. For each window all $w-k+1$ canonical $k$-mers are generated and a hash function $h_1$ is applied to each. The $s$ smallest hash values are selected as features and form the minhashing \emph{sketch}, which is used to represent the window in the database.
    Sketches are collected in batches and inserted into a different concurrent queue which is consumed by another thread. This thread is responsible for inserting the features of each sketch together with its location information (target and window) into the hash table.

    When using GPUs a consumer thread collects the sequence data and sends it to the GPU in batches. The processing and database insertion is handled on the device. In a multi-GPU environment we spawn as many consumer threads as there are GPUs, each thread scheduling work on a distinct GPU.
    Note, that a single reference sequence will never be distributed across multiple GPUs. However, the same $k$-mer might be present in several hash tables because it can appear in multiple genomes. This partitioning is deliberate and helps to reduce data communication in the classification phase.

    In the CPU version the hash table uses open addressing where each slot maps a feature to a bucket of reference locations. To counteract the biased distribution of sketch values a second hash function $h_2$ is applied in order to determine the key slot in the hash table. If the computed key slot is empty a new bucket is created for the feature and the value is inserted. If the feature is already present in the slot the value is appended to the corresponding bucket. If the slot is occupied by a different feature a quadratic probing scheme is used to find the next slot where inserting is tried again.
    Since the distribution of locations per $k$-mer is usually highly skewed (a large fraction of $k$-mers occur only once while few occur many times) the buckets can grow dynamically using a geometric growth scheme. Additionally, the maximum number of locations stored per $k$-mer is limited to a pre-defined value ($254$ per default).

    Although multiple threads are used in this pipeline, MetaCache's default hash table does not support concurrent insertion. It uses a dynamic allocation strategy which allows it to grow if the load factor exceeds a user defined limit. In this case a new hash table of larger size is allocated into which all feature-bucket mappings are re-inserted while the buckets holding the values are preserved. Therefore, the CPU version of MetaCache is limited to a single thread operating the hash table in the build phase. However, the benefit is, that the locations in each bucket remain sorted throughout the whole process due to the ascending target and window IDs assigned by the sketching thread. This allows linear-time merging of multiple location lists into one sorted list, as is required in the query phase.

    After database construction has finished, the taxonomic meta information as well as the hash table are written to the file system.

\subsection{Query Phase} \label{sec:cpu:query}

    If MetaCache is not executed in on-the-fly query mode following the build phase, the database has to be read from disk before reads can be queried. Here a condensed form of the hash table is used where all buckets of target locations are loaded into one large contiguous array which can be accessed by pointers. The on-the-fly mode uses the hash table from the build phase as is. In addition, an acceleration structure is generated from the taxonomic tree that contains the taxonomic lineage of each target in the database thus allowing to compute the lowest common ancestor of two taxa in constant time during classification.

    Similar to the build phase, a concurrent queue is used to distribute work among multiple threads.
    One thread is responsible for reading sequences from input files and separating header from sequence data. Batches of sequences are inserted into the queue. Multiple threads consume the batches from the queue and process all contained reads. To classify a read, its sequence is first split into windows of the same length as used in the database. From each window all canonical $k$-mers are generated and minhashing is applied to produce a sketch. All elements of the sketch are then queried against the hash table. The resulting location lists are merged and identical locations are accumulated. This yields a (sparse) histogram of hit counts per window in the reference genomes ({\em window count statistic}) which indicate the similarity of this region with the read.

    To account for single-end or paired-end reads spanning multiple windows the window count statistic is scanned with a sliding window approach to find target regions with the highest aggregated hit counts in a contiguous window range. The top $m$ counts (\emph{top hits}) are then used to classify the read. Usually $2\leq m \leq 4$ top hits are enough to achieve a reliable classification. If the difference of the highest and second highest count is above a threshold, the read is labeled as belonging to the taxon of the genome corresponding to the maximum count. Otherwise, all targets with counts close to the maximum are considered, the lowest common ancestor of the corresponding taxa is calculated and used to label the read.

    In the GPU version the consumer threads split reads into windows and send batches of these sequence windows to the first GPU. Because the database may be distributed across multiple GPUs, each device has to be queried to get the classification result.
    After generating sketches on the first GPU, they are sent to the other GPUs. Each GPU generates its own top hits for each query, which are then sent to the next GPU and merged with its local top hits. We finally obtain the top hits of the whole database on the last GPU. This result is copied back to the host, which assigns the final classification for each read.

\subsection{Database Partitioning}

    Due to the ever increasing size of reference genome collections, the memory of a single workstation might not be sufficient for MetaCache to create a database for all reference sequences at once.
    Therefore, it is possible to partition the references genomes into separate databases which allows to save memory either while building or while querying or during both phases.

    Partitioning of the reference genomes can be done as a preprocessing step followed by successive distinct builds thus minimizing memory usage. Alternatively, if the workstation used for building has enough resources, MetaCache can also build database partitions in parallel.

    Partitioned databases can be queried sequentially using independent query runs followed by a merge step to obtain the final classification result. Alternatively, all or a subset of partitions can be loaded at the same time and be queried in parallel. Furthermore, partitioned databases generated by the CPU version of MetaCache can be loaded by the GPU version and vice versa as long as no database part exceeds the memory of a single GPU.

\section{MetaCache GPU Implementation}\label{sec:gpu:impl}
%================================================================================

    The overall structure of MetaCache is the same for the CPU and GPU versions. However, many data structures and sub-routines of the program had to be adapted to run efficiently on the GPU. We employ a novel, specialized hash table on the GPU to accelerate database building, querying, and classification. To overcome the memory limitations of a single GPU we extend MetaCache to work with hash tables distributed across multiple GPUs.
    Figure \ref{fig:gpu_pipeline} shows an overview of the GPU workflow.

\subsection{GPU Hash Tables} \label{sec:gpu:build}

    When building a database with GPUs, the hash tables are allocated in GPU memory, while meta information like the taxonomic tree remain in host memory. For the hash tables we use a static allocation strategy to avoid costly resizing of the data structure which would stall the parallelized insertion process.

    \begin{figure}[t]
        \centering
        \includegraphics[width=.66 \linewidth]{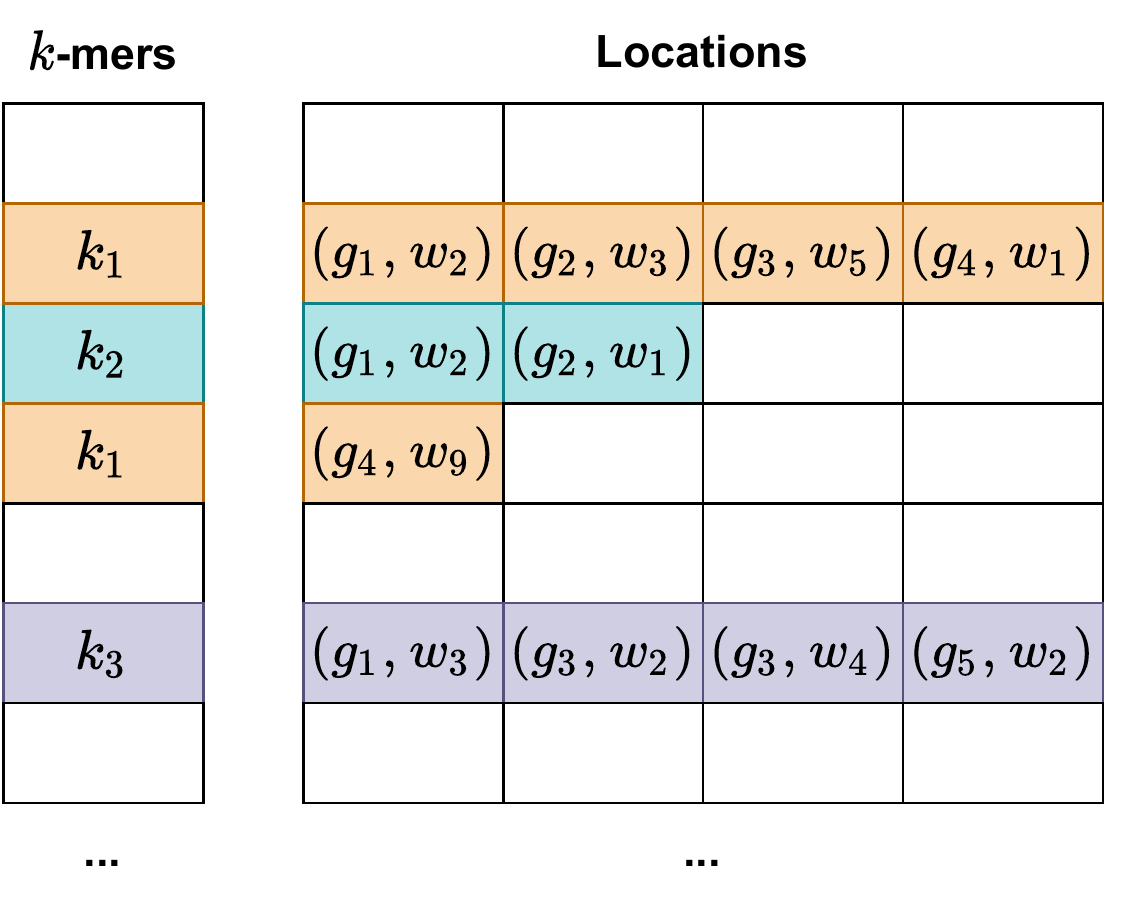}
        \caption{Multi Bucket Hash Table Layout. Each slot maps a $k$-mer to a fixed number of locations.
        % Note that multiple slots may belong to the same $k$-mer.
        }
        \label{fig:multi_bucket_hash_table}
    \end{figure}

    Using WarpCore we designed and implemented a new open addressing hash table variant, where each slot consists of a key mapped to a small, fixed number of values. The layout of this \emph{Multi Bucket Hash Table} is illustrated in Figure \ref{fig:multi_bucket_hash_table}. Here, a key can occur in multiple slots which allows it to be associated to an arbitrary number of values. Compared to WarpCores's \emph{Multi Value Hash Table} where each slot accommodates only a single key-value pair and WarpCores's \emph{Bucket List Hash Table} where each key is mapped to a linked list of buckets, this new variant is a better fit to the various key-value distributions that we encountered in our experiments. It consumes less memory than the others, which conversely allows for more data to be stored per GPU.

    % Note, that WarpCores's Bucket List Hash Table was able to achieve slightly greater insertion speed than the Multi Bucket Hash Table, but besides the memory overhead it comes with the additional disadvantage that the available GPU memory needs to be divided into space for key and value data so their ratio needs to be known or estimated beforehand. The Multi Bucket Hash Table does not have this problem and can accommodate different key-value distributions without changing the configuration.

    In case the query is performed using an already existing database, MetaCache loads the data from files into the GPUs using a different structure. Similar to the CPU version, we use a condensed layout where all buckets are stored in one contiguous array in GPU memory. To associate keys and buckets we employ WarpCore's \emph{Single Value Hash Table} to map the keys to pointers into the array.
    Note that if the query phase is started directly after a build, the hash table is used as-is and will not be compacted.
    % Additionally, we copy the acceleration structure containing the taxonomic lineages to each GPU.

\subsection{GPU Pipeline}

    To enable multiple host threads to provide work while limiting memory occupancy on the devices, we use a pipeline approach, allocating memory for all steps needed for processing a single batch of sequences on each GPU. CUDA events are used to orchestrate the pipeline, signaling when a stream has to wait or can continue work using the same memory resources as its predecessor.

    The query pipeline on the GPU consists of the following steps:
    \begin{enumerate}
        \item Encode all sequence characters of a window into 3-bit representations of the nucleotide bases (A,C,G,T,N). \label{sec:gpu:pipeline:encode}
        \item Generate all valid $k$-mers and hash them using function $h_1$.
        \item Sort the hashes to get the minhashing sketch.
        \item The sketch is queried against the hash table and resulting locations are written to memory. \label{sec:gpu:pipeline:query}
        \item Compact the location lists of all processed windows.
        \item Execute a segmented sort to sort the locations for each read. \label{sec:gpu:pipeline:sort}
        \item Obtain the window count statistic by accumulating identical locations. \label{sec:gpu:pipeline:accum}
        \item Find the top hits using a sliding windows scan. \label{sec:gpu:pipeline:slide}
    \end{enumerate}

    In contrast, the build pipeline only performs the first three steps, and then inserts the sketch with the corresponding target information into the database.

    Our GPU kernels combine multiple steps of the pipeline and have been optimized for efficient work sharing between many CUDA threads. We employ groups of 32 threads (so-called \emph{warps}) to tackle the same problem. We enforce additional constraints on MetaCache's sub-sampling parameters to improve data layout and memory access patterns. Especially, we require the offset between window beginnings (\emph{window stride}) to be a multiple of $4$ to enable aligned access to $4$ characters per thread.
    Note that the default parameters are $k$-mer length of $k=16$ characters, a sketch size of $s=16$, a window length of $w=127$ characters and a window overlap of $k-1$ which results in a window stride of $127-16+1 = 112$.
    Kernel implementation details are presented in the following.

    % Note, that an almost identical kernel is used for inserting values in the build phase.

\subsection{Minhashing and Querying}

% 	\begin{figure*}[t!]
% 	    \includegraphics[width=0.9\textwidth]{figures/kmer_generation.png}
% 		\caption{example of $k$-mer generation using a warp of size 12: (1) Load 4 characters per thread (yellow). (2) Get characters of sub-warp (orange). (3) Get characters from subsequent sub-warp (green). (4) Generate four $16$-mers from framed characters.}
% 		\label{fig:kmer_generation}
% 	\end{figure*}

    The first kernel executes steps (\ref{sec:gpu:pipeline:encode})--(\ref{sec:gpu:pipeline:query}) in the GPU pipeline. Each warp processes the sequence characters of a single window of maximum size $128$ at a time. First, each thread in the warp loads $4$ consecutive characters from global memory into register using a $4$-byte load operation. Next, it encodes the characters using a 2-bit representation of the regular nucleotide bases A,C,G,T and combines them in a single 32-bit integer. Ambiguous base characters N are noted as single bits in an auxiliary integer. 4 adjacent threads form a sub-warp and combine their integers using XOR shuffle operations so that every thread holds the information of $4\cdot 4 = 16$ consecutive characters. Then each sub-warp gets the information from the subsequent sub-warp by means of another shuffle operation. Now each thread contains the data of $32$ characters, overlapping by $16$ characters with the next sub-warp. From these characters thread $i$ is able to generate four $k$-mers starting at positions $4i,\dots,4i+3$ of the window, which are then hashed using hash function $h_1$.
    % The data sharing between threads of a warp is visualized in Figure \ref{fig:kmer_generation}.
    Note, that the last few threads (number depends on window and $k$-mer length) do not generate any $k$-mers because they would exceed the window boundary.

    To produce the minhashing sketch we have to find the $s$ smallest hash values in the warp. We reorder all hash values using a bitonic sort implementation similar to Section \ref{sec:gpu:impl:sort} which operates only on registers with the help of warp shuffles. Next we remove duplicates and store the $s$ smallest unique values in shared memory.

    Subsequently, all features of the sketch are queried against the hash table.
    WarpCore provides device-sided hash table operations which can be called from inside the kernel. It uses a cooperative probing scheme which means that multiple threads work together as a group to insert or find a key. Therefore we split our warp into groups to retrieve the bucket pointers to the locations associated with the queried keys. After replacing the sketch with these pointers in shared memory, all threads in the warp work in unison again to retrieve the target locations from each bucket. The locations are then stored in global memory.
    %In case of an odd number of total locations for this window, we pad the list with a dummy value at the end in order to achieve better memory alignment.

\subsection{Compaction}

    The next step is preceded by a prefix sum over the number of target locations per window. With this information we are able to compact the location lists of the whole batch of windows. The compaction kernel uses one thread block per window to copy the locations from the result array from the first kernel to a dense array. The induced alignment allows each thread to efficiently copy two locations at once, which increases throughput. The kernel also checks if consecutive windows originate from the same read to calculate the segment boundaries needed for the sorting step.

\subsection{Segmented Sort} \label{sec:gpu:impl:sort}

    Step (\ref{sec:gpu:pipeline:sort}) of the pipeline sorts the location list for each read. Our segmented sort implementation is a highly modified key-only version of \cite{Segsort}. Our version reduces the memory overhead while being able to run asynchronously with respect to the host thread. Additionally, we bundle most of the kernels used in this step in a CUDA graph which speeds up the scheduling and removes gaps between kernel executions.

\subsection{Top Candidate Generation}

    The last kernel combines steps (\ref{sec:gpu:pipeline:accum}) and (\ref{sec:gpu:pipeline:slide}).
    Each warp running the kernel is used to find the top hits of one read at a time.
    The sliding window size $sws$ is determined by the length of the respective read and defines the maximum number of contiguous locations that can be accumulated to form a top candidate.
    First, threads load the sorted locations and perform a segmented reduction to accumulate identical values. This is repeated until at least $32+sws-1$ unique locations are collected in shared memory, so that every thread is able to calculate its own sliding window. Next, each thread has to inspect up to $sws$ locations to determine if they belong to the same region by comparing their target and window ids. Starting with its first locations each thread either accumulates the scores of consecutive locations or discards all following locations if their ids are out of range. The resulting location ranges and their scores are potential top candidates.

    Because the number $m$ of top candidates required to classify a read is small, each thread is able to maintain its own list of $m$ ranges with highest hit count in registers. After each iteration of the candidate generation the local top lists are updated individually. Finally, after all locations for a read have been processed the whole warp generates the combined top hits list by using warp shuffles to find the highest scores.

%================================================================================
\section{Performance Evaluation} \label{sec:evaluation}
%================================================================================

    To evaluate the performance of our approach we investigated the build and query phases for two different databases of reference genomes. The first set consists of genomes from NCBI RefSeq     \cite{o2016reference}. We included all complete archaeal, bacterial, fungal and viral genomes from RefSeq Release 202. For the second set we combined the RefSeq set with 31 food-related genomes from the All-Food-Sequencing (AFS) pipeline \cite{kobus2020big}. In contrast to the first set these animal and plant genomes consist of much longer sequences. Furthermore, some of the genomes are only available at scaffold level which results in hundreds of thousands of different target sequences per genome.
    Table \ref{tab:databases} shows the number of included species for each database as well as their total sizes.

    \begin{table}[t]
        \centering
        \caption{Reference genome sets used for databases.
        %RefSeq 202 includes archaeal, bacterial, fungal and viral genomes. All-Food-Seq (AFS) 31 consists of animal and plant genomes relevant for food stuff.
        }
        \begin{tabular}{lrr}
            \toprule
            Database             & Species      & Size on disk \\
            \midrule
            RefSeq 202           & 15,461       & 74 GB \\
            All-Food-Seq 31      &     31       & 77 GB \\
            \midrule
            AFS 31 + RefSeq 202  & 15,492       & 151 GB \\
            \bottomrule
            % \toprule
            % Database             & References & Size on disk \\
            % \midrule
            % RefSeq 202           & 29,240      & 74 GB \\
            % All-Food-Seq 31      &     31      & 77 GB \\
            % \midrule
            % AFS 31 + RefSeq 202  & 29,271      & 151 GB \\
            % \bottomrule
        \end{tabular}
        \label{tab:databases}
    \end{table}

    \begin{table}[t]
        \caption{Metagenomic read datasets.}
        \centering
        \begin{tabular}{lcrrrr}
            \toprule
            Dataset & Format       & Sequences  & Min & Max & Average \\
            \midrule
            HiSeq   & FASTA single & 10,000,000 & 19  & 101 & 92.3 \\
            MiSeq   & FASTA single & 10,000,000 & 19  & 251 & 156.8 \\
            KAL\_D  & FASTQ paired & 26,114,376 & 101 & 101 & 101 \\
            \bottomrule
        \end{tabular}
        \label{tab:datasets}
    \end{table}

    For the query phase we chose to test three metagenomic datasets with at least 10 million reads each.
    HiSeq and MiSeq contain single-end reads in FASTA format and were introduced by Wood and Salzberg \cite{kraken}, KAL\_D is taken from AFS \cite{kobus2020big} and contains paired-end reads in the FASTQ fromat. Table \ref{tab:datasets} shows the total number of reads per dataset together with their minimum, maximum and average lengths.
    The datasets HiSeq and MiSeq represent bacterial mock communities consisting of reads from ten different species each, produced by Illumina sequencers of the same names.
    Finally, KAL\_D is a real world dataset obtained by sequencing material from a sausage made from beef, mutton, pork and horsemeat.

    Experiments were conducted on the following system:
    \begin{description}
        \item[DGX-1 Volta:] Dual-socket Xeon E5-2698 v4 CPU (2x20 cores at 2.20 GHz) with 512 GB DDR4 RAM and 8 Tesla V100 GPUs, each with 32 GB HBM2 memory, CUDA 11.0, GCC 9.3.0.
    \end{description}

    Reference genomes, taxonomy files and datasets were loaded into a virtual RAM drive before executing a build or query to minimize the influence of slow I/O from the file system. Kraken2 and MetaCache (CPU version) were executed using a maximum of 80 threads incorporating simultaneous multithreading on the 40 cores of the system.
    MetaCache-GPU was evaluated using a 4 GPU and an 8 GPU configuration.
    In the 4 GPU configuration our Multi Bucket Hash Table needed 10\% and 11\% less memory than WarpCore's Multi Value and Bucket List Hash Table, respectively. It was the only hash table that could fit RefSeq202 on 4 GPUs without further restricting the number of locations per $k$-mer.
    The larger AFS31+RefSeq202 database did not fit in the memory of 4 V100 GPUs and therefore always uses 8 GPUs.

\subsection{Build Performance}

    \begin{table}[t]
        \caption{Build performance for different databases. Total time includes build time and time for writing DBs to files.}
        \centering
        \begin{tabular}{lcrrrrr}
            \toprule
            Method & Build time & Total time & DB size & RAM \\
            \midrule
            % \multirow{4}{*}{RefSeq 202}
            \multicolumn{4}{l}{RefSeq 202 database:} \\
            \midrule
            Kraken2   & -      & 72 min & 40 GB & 46 GB \\
            MC CPU    & 67 min & 69 min & 51 GB & 71 GB\\
            MC 4 GPUs & 10.4 s & 59.6 s & 88 GB & 1 GB \\
            MC 8 GPUs & 9.7 s  & 67.0 s & 97 GB & 1 GB \\
            \midrule
            % \multirow{3}{*}{AFS 31 + RefSeq 202}
            \multicolumn{4}{l}{AFS 31 + RefSeq 202 database:} \\
            \midrule
            Kraken2   & -          & 256 min & 110 GB & 160 GB \\
            MC CPU    & 194 min & 201 min & 127 GB & 194 GB \\
            MC 8 GPUs & 42.7 s     & 3 min 31 s & 176 GB & 30 GB\\
            \bottomrule
        \end{tabular}
        \label{tab:build}
    \end{table}

    Table \ref{tab:build} presents the build performance of Kraken2 and MetaCache's CPU and GPU versions.
    As mentioned in Section \ref{sec:cpu:build} the CPU version of MetaCache is based on a two stage producer-consumer scheme which uses only three threads in total. Nevertheless, it is faster than Kraken2 which uses 80 threads. While CPU-based MetaCache and Kraken2 take more than an hour for building the RefSeq202 database and more than 3 or more than 4 hours for AFS31+RefSeq202, respectively, MetaCache-GPU is able to create the index structures in seconds to minutes. The speedup when using 8 GPUs is 64x and 61x for building and storing RefSeq202 compared to Kraken2 and MetaCache-CPU, respectively. For AFS31+RefSeq202 the speedups are 72x and 51x, respectively. Looking at the build time without writing the databases to the file system, the GPU version is 414 times and 272 times faster than the CPU version of MetaCache for RefSeq202 and AFS31+RefSeq202, respectively.
    Building the RefSeq202 database using 4 GPUs is a little slower than when using 8 GPUs because of less parallelization. But the overall database size is smaller and less data needs to be written to files, resulting in a slightly faster total runtime compared to the 8 GPU version. The speedups are 72x and 69x compared to Kraken2 and MetaCache-CPU, respectively.

\subsection{Query Performance}

    Table \ref{tab:query} reveals the query performance of Kraken2 and MetaCache for the databases RefSeq202 and AFS31+RefSeq202, respectively. MetaCache's speed depends on the number of reference locations found per read, which need to be merged and analyzed to get the final classification results. While the HiSeq  dataset only contains reads which are smaller than MetaCache's window size, longer reads from MiSeq are split into two windows, resulting in more database queries and likely more retrieved locations. The KAL\_D dataset on the other hand contains mostly reads from meat components and does not register many hits against the RefSeq202 database, resulting in the fastest queries for MetaCache.
    Kraken2 reaches higher speeds than MetaCache's CPU version when querying RefSeq202 with MiSeq and HiSeq, but needs more time for KAL\_D. Since Kraken2 relies on mapping minimizers directly to taxa and does not need to process locations lists for hits in the database the speed in not affected much by the database size, Kraken2 is even able to increase the speed for the bigger AFS31+RefSeq202 database, while MetaCache's query speed takes a hit and is 6 to 14 times lower.

    % !!!!!!!!!!!!!!!!!!!!!!!!!!!!!!!!!!!!!!!!!!!!!!!!!!!!!!!!!!!!!
    % wir sollten erwähnen, dass MetaCache durch den zuvor erwähnten Unterschied prinzipiell "mehr" kann als Kraken2,
    % da wir auch die most likely locations of origin ausgeben können und Kraken2 "nur" auf Taxa mappt
    % sonst könnte einem der vorige Absatz den Eindruck vermitteln, dass Kraken2s Ansatz überlegen ist und
    % wir einen dummen Ansatz parallelisiert haben
    % Daher mein Vorschlag:

    Note that Kraken2 can only map reads to candidate taxa while MetaCache is able to map reads to the most likely locations of origin within reference sequences and thus produce candidate regions for further downstream analysis like, e.g., alignments.

    MetaCache's GPU version however is not affected by the database size and achieves high speeds for both RefSeq202 and AFS31+ RefSeq202. It is able to outperform Kraken2 and MetaCache's CPU version on all datasets. Compared to Kraken2 MetaCache-GPU is 1.9-6.2 times faster for RefSeq202 and 2-3.3 times faster for AFS31+RefSeq202, compared to MetaCache-CPU it is 2.8-11.2 and 23-153 times faster, respectively.

    \begin{table}[t]
        \centering
        \caption{Query performance against different database. Query speed in million reads per minute.}
        \begin{tabular}{llcrrrr}
            \toprule
            Method & Dataset & \multicolumn{2}{c}{RefSeq 202} & \multicolumn{2}{c}{AFS31+RefSeq202} \\
            && Time & Speed & Time & Speed \\
            \midrule
            \multirow{3}{*}{Kraken2}
            & HiSeq	 &  4.6 s & 130  &  4.1 s &  147   \\
            & MiSeq	 &  6.9 s &  87  &  5.9 s &  102   \\
            & KAL\_D & 42.6 s &  74  & 37.8 s &   83   \\
            \midrule
            \multirow{3}{*}{MC CPU}
            & HiSeq	 & 11.4 s & 53  & 107.4 s & 5.6  \\
            & MiSeq	 & 31.2 s & 19  & 464.6 s & 1.3  \\
            & KAL\_D & 38.9 s & 81  & 237.0 s & 13  \\
            \midrule
            \multirow{3}{*}{MC 4 GPUs}
            & HiSeq	 &  2.1 s & 292  & --\footnotemark & --\footnotemark[\value{footnote}] \\
            & MiSeq	 &  3.6 s & 165  & --\footnotemark[\value{footnote}] & --\footnotemark[\value{footnote}] 	\\
            & KAL\_D &  6.9 s & 454  & --\footnotemark[\value{footnote}] & --\footnotemark[\value{footnote}] 	\\
            \midrule
            \multirow{3}{*}{MC 8 GPUs}
            & HiSeq	 &	2.0 s	& 305  &  2.0 s & 298  \\
            & MiSeq	 &	2.8 s	& 215  &  3.0 s & 199  \\
            & KAL\_D &	7.2 s	& 435  & 12.6 s & 249  \\
            \bottomrule
        \end{tabular}
        \label{tab:query}
    \end{table}
    \footnotetext{4 V100 GPUs do not provide enough memory for AFS31+RefSeq202.}

\subsection{On-The-Fly Mode}

\begin{figure}[t]
        \centering
        \includegraphics[width=\linewidth]{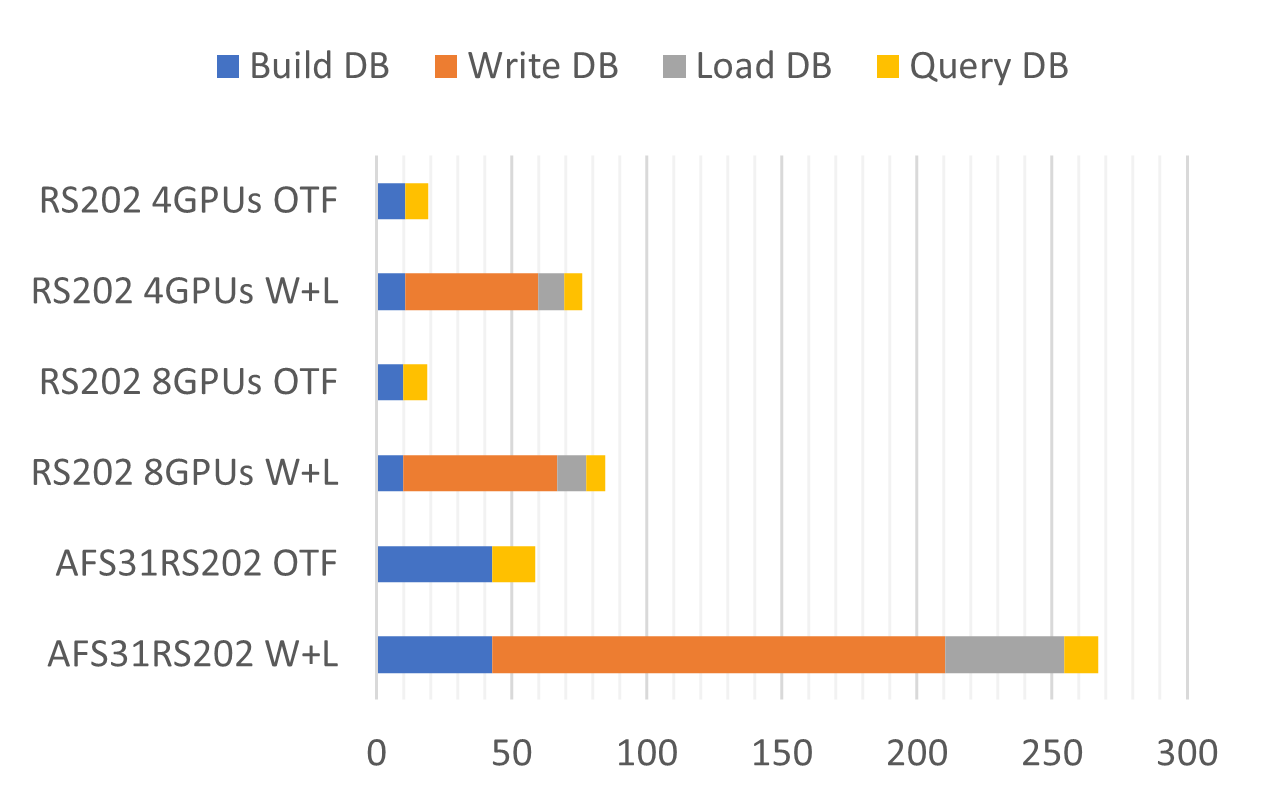}
        \caption{Runtime comparison of our on-the-fly (OTF) mode to separate build and query execution (W+L) for different databases, querying KAL\_D dataset.}
        \label{fig:on-the-fly}
    \end{figure}

    Figure \ref{fig:on-the-fly} compares the runtimes of separate build and query to the on-the-fly mode mentioned in Section \ref{sec:cpu} using multiple GPUs. Because of the high GPU build speed, most of the time in the build phase is actually spent writing the database to the file system. Loading the database takes almost the same time as building it from scratch. When using the on-the-fly mode the database can be queried without having to write and reload the database.
    Table \ref{tab:otf} shows that the time needed until a query can be executed is greatly reduced when using MetaCache-GPU in on-the-fly mode. MetaCache's GPU databases are ready for use in under a minute which translates to a speedup of 360-450 compared to Kraken2.
    Note, that the query speed of the GPU hash table used for building is lower than that of the one used in separate query execution resulting in about 20\% less performance. Nevertheless the on-the-fly query mode can be beneficial in situations where a database is not queried more than once and does not need to persist on disk.

    \begin{table}[t]
        \caption{Comparison of time needed until a query can be executed when using MetaCache's on-the-fly (OTF) mode. TTQ is Time-to-Query.}
        \centering
        \begin{tabular}{lcrrrr}
            \toprule
            Method & Build & Load & TTQ & Speedup \\
            \midrule
            % \multirow{4}{*}{RefSeq 202}
            \multicolumn{4}{l}{RefSeq 202 database:}\\
            \midrule
            Kraken2       & 72 min & 23 s & 73 min & 1.0 \\
            MC CPU OTF    & 67 min & -    & 67 min  & 1.1 \\
            MC 4 GPUs OTF & 10.4 s & -    & 10.4 s     & 420 \\
            MC 8 GPUs OTF &  9.7 s & -    &  9.7 s     & 450 \\
            \midrule
            % \multirow{3}{*}{AFS 31 + RefSeq 202}
            \multicolumn{4}{l}{AFS 31 + RefSeq 202 database:}\\
            \midrule
            Kraken2       & 256 min & 63 s & 257 min & 1.0 \\
            MC CPU OTF    & 201 min & -    & 201 min & 1.3 \\
            MC 8 GPUs OTF & 42.7 s  & -    & 42.7 s  & 360 \\
            \bottomrule
        \end{tabular}
        \label{tab:otf}
    \end{table}

\subsection{Performance Breakdown}

\begin{figure}[t]
        \centering
        \includegraphics[width=\linewidth]{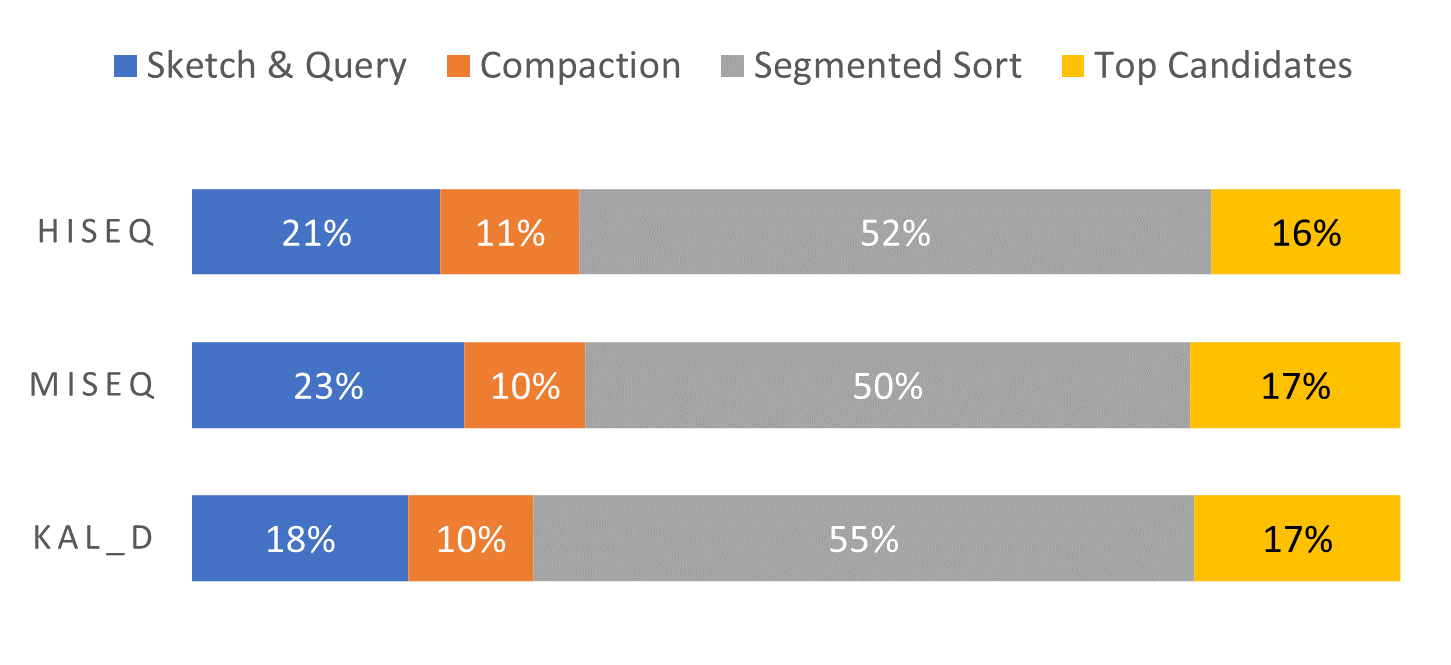}
        \caption{Performance breakdown for GPU queries against AFS31+RefSeq202 database.}
        \label{fig:performance_breakdown}
    \end{figure}

    Figure \ref{fig:performance_breakdown} illustrates the shares in the total runtime of the components of the query pipeline explained in Section \ref{sec:gpu:impl} when querying different datasets against the AFS31+RefSeq202 database. Creating the minhashing sketch from the reads and querying the database takes 18-23\% of the time while the rest is spent on processing the retrieved location lists. Segmented sort takes the biggest share of the pipeline and is responsible for about halve of the total runtime.

\subsection{Query Accuracy}

    To analyze the query accuracy we compared classification results for HiSeq and MiSeq from MetaCache's versions and Kraken2 to the ground truth.
    Table \ref{tab:query:accuracy} reveals the precision and accuracy using the RefSeq202 database with different methods.
    On the genus level Kraken2 offers greater sensitivity for the HiSeq datasets and similar sensitivity for MiSeq. The genus-level precision over 99\% is comparable for Kraken2 and MetaCache with a small advantage for Kraken2 regarding MiSeq and MetaCache for the other two datasets.
    MetaCache is able to surpass Kraken2's accuracy on the species level for HiSeq and MiSeq, yielding 5\% and 12\% more sensitivity, respectively. MetaCache's precision is also higher for HiSeq, but lower for MiSeq.

    Compared to MetaCache's CPU version the GPU variants are able to improve the accuracy. The reason for this is that using multiple database parts allows to store more locations for each $k$-mer, which are lost in the CPU version due to the enforced bucket limit. The additional location information leads to a better sensitivity and precision in most cases, only species-level results for HiSeq are slightly worse. This effect increases when using more GPUs.

    Note, that we only tested Kraken2 and MetaCache with default parameters. Both algorithms allow the user to choose a different hit threshold which defines how many database hits are necessary to classify a read. Lowering the threshold typically trades precision for sensitivity while an increased threshold may improve precision at the cost of sensitivity.

    For the KAL\_D dataset there is no true per-read mapping available, only the ratio of meat components is known. To examine this dataset we queried the AFS31+RefSeq202 database which includes the corresponding genomes.
    Using MetaCache's abundance estimation functionality achieved quantification results close to the true ratios with a accumulated deviation of 6.5\% and 2.5\% false positives for the GPU version and a deviation of 16.0\% and 2.0\% false positives for the CPU version. In contrast, comparing the species results from Kraken2's sample report to the truth yielded a deviation of 21.4\% and 7.5\% false positives.

    \begin{table}[t]
        \centering
        \caption{Classification accuracy using RefSeq 202 database.}
        \begin{tabular}{lccccc}
            \toprule
            Dataset & Method & \multicolumn{2}{c}{Species} & \multicolumn{2}{c}{Genus}  \\
                    &        & Prec. & Sens.     & Prec. & Sens. \\
            \midrule
            \multirow{4}{*}{HiSeq}  & Kraken2   & 82.52\% & 58.39\% & 99.09\% & 88.46\% \\
                                    & MC CPU    & \textbf{89.41}\% & \textbf{63.68}\% & 99.20\% & 81.36\%\\
                                    & MC 4 GPUs & 88.70\% & 62.61\% & \textbf{99.36}\% & 82.32\%\\
                                    & MC 8 GPUs & 88.81\% & 62.63\% & \textbf{99.36}\% & \textbf{82.40}\%\\
                                    %& MC CPU M4 & 88.79\% & 65.60\% & 99.05\% & 84.05\%\\
                                    %& MC 8 GPUs M4 & 88.17\% & 64.48\% & 99.21\% & 84.93\%\\
            \midrule
            \multirow{4}{*}{MiSeq}  & Kraken2   & \textbf{77.91}\% & 48.53\% & \textbf{99.38}\% & 93.25\% \\
                                    & MC CPU    & 72.28\% & 60.67\% & 99.21\% & 93.23\%\\
                                    & MC 4 GPUs & 73.07\% & 61.55\% & 99.37\% & 93.82\%\\
                                    & MC 8 GPUs & 73.53\% & \textbf{61.99}\% & 99.37\% & \textbf{93.92}\%\\
                                    %& MC CPU M4 & 71.79\% & 61.20\% & 99.16\% & 94.11\%\\
                                    %& MC 8 GPUs M4 & 73.13\% & 62.49\% & 99.32\% & 94.73\%\\
            \bottomrule
        \end{tabular}
        \label{tab:query:accuracy}
    \end{table}

%================================================================================
\section{Conclusion}\label{sec:conclusion}
%================================================================================
The steadily increasing amount of available reference genomes and NGS data establishes the need for efficient and highly optimized processing approaches. In this paper we have presented MetaCache-GPU -- an alignment-free method for metagenomic read classification on CUDA-enabled GPUs based on massively parallel construction and querying of a novel hash table structure for $k$-mers. MetaCache-GPU's on-the-fly mode enables classification pipelines that can be rapidly updated to make use of the latest reference genomes or use custom reference genome sets on demand achieving over two orders-of-magnitude speedup compared to Kraken2 and the CPU version of MetaCache while still being memory-efficient. This is particular important for the analysis of complex biological matters such as food stuff which often requires custom reference databases where the size of individual genomes can exceed several gigabytes (e.g., plant genomes).

Our current implementation takes advantage of multiple GPUs within the same node to process large-scale metagenomic databases in memory. As part of our future work we plan to extend our method to use even more GPUs within cluster systems.
%mostly infeasible to always include all available reference genomes in one database given the currently available amounts of main memory

%Also, large-scale surveys involving several hundred gigabyte-sized genomes will most likely have to split up reference databases.

Note that the construction and querying of $k$-mer indices is a common concept in bioinformatics. Thus, the introduced methods could easily be adapted to related NGS tasks such as read mapping or long-read-to-long-read alignment, which is part of our future work. Furthermore, it would also be interesting to investigate extensions that allow for scaling to larger number of GPUs on cluster systems.

% trade of sensitivity for speed by increasing the window length

% weil wir so superschnell sind, ist long read mapping so richtig toll, muss man aber aufpassen, denn je länger die reads, desto mehr windows;
%Robin - in der GPU version hattest du die window length auf 127 festgelegt? oder? dann sollte man es eher nicht erwähnen

%Ein Satz von wegen leichter parameter-anpassung:
%hit threshold allows to trade off precision for sensitivity

%Ein Satz zu mehr output möglich als andere tools:
%output of locations within the candidate reference genomes for further analysis, e.g., alignments
%built-in abundance analysis

%what about scaling to more GPUs? can we show some kind of weak scaling (more ref. genomes) - we only have 4GPUs vs. 8, right?
%but in principle scalable to more without much performance deterioration

We plan to make MetaCache-GPU publicly available upon acceptance of this paper.

%================================================================================

\begin{acks}
\begin{anonsuppress}
We acknowledge funding by the DFG project HySim. Parts of this research were conducted using the supercomputer Mogon II which is a member of the AHRP and the Gauss Alliance e.V.
\end{anonsuppress}
\end{acks}

\bibliographystyle{ACM-Reference-Format}
\bibliography{main_icpp21}

% \appendix

\end{document}